\documentclass[journal]{IEEEtran}
\ifdefined\pdfminorversion\pdfminorversion=7\fi
\usepackage{amsmath,amssymb,bm}
\usepackage{graphicx}
\usepackage[table]{xcolor}
\usepackage{cite}
\usepackage{url}
\usepackage{algorithm}
\usepackage{algorithmic}
\usepackage[caption=false,font=footnotesize]{subfig}
\makeatletter
\let\ftype@algorithm\ftype@figure
\makeatother

\let\origref\ref
\renewcommand{\ref}[1]{\textcolor{red}{\origref{#1}}}

\newcommand{\CN}{\mathcal{CN}}
\newcommand{\Hh}{\mathrm{H}}
\newcommand{\T}{\mathrm{T}}
\newcommand{\rev}[1]{#1}
\newenvironment{revision}{\par}{\par}

\title{\huge Beyond LLM-GA: Secure Fluid Antenna Systems with ReEvo-Designed Memetic Algorithm}
\author{Hanyong Xu, Zhaolai Dang, and Tong Zhang
\thanks{This work was supported in part by the National Natural Science Foundation of China under Grant 62401229.}
\thanks{H. Xu and Z. Dang are with the School of Information, Harbin Institute of Technology, Shenzhen, Shenzhen 518055, China (e-mail: \{2025311252, 2025310700\}@stu.hit.edu.cn).}
\thanks{T. Zhang is with Guangdong Provincial Key Laboratory of Aerospace Communication and Networking Technology, Harbin Institute of Technology, Shenzhen, 518055, China (e-mail: tongzhang@hit.edu.cn).}
\thanks{Code is available at \textcolor[RGB]{135,206,235}{\protect\url{https://github.com/Xuhanyong/secure-fas-llmga}}.}
}

\begin{document}
\maketitle

\begin{abstract}
Fluid antenna systems (FASs) offer significant spatial flexibility, yet securing them against eavesdropping is critical for practical FAS deployment in military, satellite, and internet-of-things networks. Although large language model (LLM)-assisted genetic algorithms (LLM-GAs) can address this secure FAS port selection problem, whether further algorithmic improvement is possible warrants deeper investigation. To this end, we propose a memetic algorithm based on reflective evolution (ReEvo). Unlike the state-of-the-art LLM-GAs, which design only crossover or mutation operators with an LLM, our algorithm leverages an LLM to evolve dedicated crossover, mutation, and local search operators offline. These operators are then embedded into a memetic search framework, thereby obviating any online LLM queries during execution. Simulation results at equal generation counts demonstrate that our proposed algorithm achieves a higher secure sum-rate than the conventional GA and the state-of-the-art LLM-GAs. 
\end{abstract}


\section{Introduction}
The ever-increasing demand for high data rates and massive connectivity in next-generation wireless networks calls for new degrees of freedom to enhance spectral efficiency and link reliability. To this end, fluid antenna systems (FASs) enable antenna port selection over a dense aperture, improving signal strength, interference suppression, and spatial diversity without extra radio frequency (RF) chains \cite{new2025fas,peng2026acc6g,ieee11587674}. Unfortunately, joint port selection and beamforming, if optimally designed, holds the key to fully unlocking the potential of FAS, yet it poses a mixed-integer nonconvex problem that is computationally challenging \cite{newadded1,newadded2,magrpo2026}.

Securing FAS wireless communications against passive eavesdropping poses a distinct and formidable challenge. This, combined with the need to jointly optimize discrete port selections and continuous beamforming vectors, significantly complicates the design, as the secrecy objective is nonconvex and intricately coupled with multiuser interference and power limitations. Analytical studies characterize secrecy capacity, outage probability, and the impact of spatial channel asymmetry between legitimate users and eavesdroppers \cite{zou2016pls} and \cite{ghadi2024pls}. Recent works jointly optimize beamforming and antenna positions in FAS-aided non-orthogonal multiple access (NOMA) systems \cite{mai2024noma}, or use penalty and majorization--minimization procedures for secure and covert links \cite{yao2024covert}. However, none of the above works has fully addressed the secure port selection problem.

Large language models (LLMs) have recently been used to address the port selection/antenna position selection problem \cite{wang2025llmfas,deng2025llmsurvey,wang2026fas,zheng2026llmfas}. In particular, Wang \emph{et al.} \cite{wang2026fas} applied reflective
evolution (ReEvo) \cite{ye2024reevo} to a binary-genetic algorithm (GA) crossover for power minimization. Their crossover uses population fitness, elite-port frequencies, and port co-occurrences, with conventional selection, mutation, elitism, and repair. The learned crossover is generated offline, has no explicit local search stage, and is not evaluated with a secrecy objective. In addition, Zheng \emph{et al.} \cite{zheng2026llmfas} proposed an efficient LLM-GA algorithm, which combines a frequency-based crossover with an LLM-designed one-port mutation. Its exchange probability and displacement range are prescribed, after which standard GA selection, elitism, and repair are applied, and the target is nonsecure multiuser performance such as max-min signal-to-interference-plus-noise ratio (SINR). Note that \cite{wang2026fas} learns only the crossover operator, whereas \cite{zheng2026llmfas} learns only the mutation operator. Port prediction has also been investigated in \cite{zhang2025portllm}.

\begin{table}[!t]
\centering
\caption{Comparison of the LLM-designed operators in \cite{wang2026fas}, \cite{zheng2026llmfas}, and the proposed algorithm.}
\label{tab:llm-ga-comparison}
\renewcommand{\arraystretch}{1.2}
\setlength{\tabcolsep}{2pt}
\begin{tabular}{@{}p{0.25\columnwidth}p{0.23\columnwidth}p{0.23\columnwidth}p{0.23\columnwidth}@{}}
\hline
Operator class & Ref.~\cite{wang2026fas} & Ref.~\cite{zheng2026llmfas} & Proposed\\
 & (crossover only) & (mutation only) & (memetic)\\
\hline
LLM-Designed Mutation & No & Yes & Yes\\
LLM-Designed Crossover & Yes & No & Yes\\
LLM-Designed Local~Search & No & No & Yes\\
\hline
\end{tabular}
\end{table}

In this paper, we propose a ReEvo-designed memetic algorithm for secure port selection in multiuser downlink FASs that outperforms state-of-the-art LLM-GAs in terms of secure sum-rate, building on the celebrated memetic algorithm \cite{moscato1989memetic}, which outperforms the conventional GA by adding a local search operator. Each candidate port subset is scored by its achievable secure sum-rate, computed via a secure weighted minimum mean-square error (WMMSE) beamforming procedure. Under this secrecy-driven fitness, ReEvo successively evolves three complementary operators offline, namely crossover, mutation, and local search. The learned crossover exploits elite-port statistics and grid topology, the learned mutation performs feasible one-in-one-out swaps at both local and global scales, and the learned local search operator generates four complementary one-swap neighbors per offspring. All three operators are frozen after offline evolution and embedded into a memetic search framework, requiring no LLM queries during online execution. Table~\ref{tab:llm-ga-comparison} contrasts the LLM-designed operators of \cite{wang2026fas} and \cite{zheng2026llmfas} with those of the proposed algorithm, whereas unlike prior LLM-GAs that design only crossover or mutation operators without security consideration, our algorithm accounts for eavesdropper leakage and performs local refinement under a secrecy-aware fitness. Simulation results demonstrate that our proposed algorithm achieves a higher secure sum-rate than the conventional GA and the state-of-the-art LLM-GAs.

\section{System Model and Problem Formulation}

%We first define the system model, and then formulate the joint secure port selection and beamforming Problem.

\subsection{System Model}
We consider a multiuser downlink FAS in which a base station (BS) serves $K$ single-antenna users in the presence of a passive single-antenna eavesdropper. Let $\mathcal K\triangleq\{1,\ldots,K\}$ denote the user index set. The BS has a rectangular FAS aperture of dimensions $\lambda W_x\times\lambda W_y$, where $\lambda=c/f_c$ and $W_x,W_y$ are normalized by the carrier wavelength. A uniform $N_x\times N_y$ grid provides $N_s=N_xN_y$ candidate ports, of which exactly $n_s$ are active.

Let $(n_{x,i},n_{y,i})$ denote the two-dimensional grid index of port $i$, where $1\leq n_{x,i}\leq N_x$, $1\leq n_{y,i}\leq N_y$, and $i=n_{y,i}+(n_{x,i}-1)N_y$. The physical distance between ports $i$ and $j$ is $d_{i,j}=\sqrt{\left(\frac{n_{x,i}-n_{x,j}}{N_x-1}\lambda W_x\right)^2+\left(\frac{n_{y,i}-n_{y,j}}{N_y-1}\lambda W_y\right)^2}$. Consistent with the channel generator, define $J_{i,j}=j_0(2\pi d_{i,j}/\lambda)$, where $j_0(x)=\sin(x)/x$ and $j_0(0)=1$. The spatial correlation matrix is given by
\begin{equation}
 \bm J=\begin{bmatrix}
 J_{1,1}&J_{1,2}&\cdots&J_{1,N_s}\\
 J_{2,1}&J_{2,2}&\cdots&J_{2,N_s}\\
 \vdots&\vdots&\ddots&\vdots\\
 J_{N_s,1}&J_{N_s,2}&\cdots&J_{N_s,N_s}
 \end{bmatrix}\in\mathbb R^{N_s\times N_s}.
\end{equation}
Because $\bm J$ is real and symmetric, write its eigendecomposition as
$\bm J=\bm U\bm\Lambda\bm U^{\T}$, where $\bm U$ is real and orthogonal.
Let $r_k$ and $r_e$ denote the BS-to-user and BS-to-eavesdropper distances,
respectively. We model the large-scale fading by
$\beta_k=\beta_0(r_k/r_0)^{-\alpha}$ for $k\in\mathcal K$ and
$\beta_e=\beta_0(r_e/r_0)^{-\alpha}$, where $\beta_0$ is the reference gain
at distance $r_0$ and $\alpha$ is the
path-loss exponent.  The channel vector of user $k\in\mathcal K$ is given by
\begin{equation}
\bm h_k^{\Hh}=\sqrt{\beta_k}\,\bm g_k^{\Hh}\bm\Lambda^{1/2}\bm U^{\T},\quad k\in\mathcal K,
\end{equation}
where $\bm g_k\sim\CN(\bm 0,\bm I_{N_s})$ for $k\in\mathcal K$.
The eavesdropper channel is given by
\begin{equation}
\bm h_e^{\Hh}=\sqrt{\beta_e}\,\bm g_e^{\Hh}\bm\Lambda^{1/2}\bm U^{\T}
\end{equation}
where $\bm g_e\sim\CN(\bm 0,\bm I_{N_s})$. Independent Gaussian vectors are
drawn for the users and the eavesdropper, and are mapped through the same correlation
matrix. Hence, the links share spatial correlation but have independent
small-scale fading. Shadowing is omitted. As an upper-bound benchmark,
we assume that the BS has perfect instantaneous channel state information
(CSI) for every link. Eavesdropper CSI may have been acquired while that
terminal was previously active.

Let $\bm e_i$ denote the $i$th column of $\bm I_{N_s}$. When $n_s$
distinct ports with indices $m_1,\ldots,m_{n_s}$ are activated, their port selection matrix is defined as $\bm E_{n_s}\triangleq
[\bm e_{m_1},\ldots,\bm e_{m_{n_s}}]\in\mathbb R^{N_s\times n_s}$.
Each column of $\bm E_{n_s}$ selects one physical port, so feasibility
requires
\begin{equation}
    \bm E_{n_s}(:,\ell)\in\{\bm e_1,\ldots,\bm e_{N_s}\},
    \quad \ell=1,\ldots,n_s.\label{eq:port-membership}
\end{equation}
and the selected ports must be distinct:
\begin{equation}
\bm E_{n_s}(:,\ell)\ne\bm E_{n_s}(:,q),\quad \ell\ne q.
\label{eq:port-distinct}
\end{equation}
Together, these constraints activate exactly $n_s$ distinct ports.
Define the effective channel vectors $\bar{\bm h}_k=\bm E_{n_s}^{\Hh}\bm h_k$
and $\bar{\bm h}_e=\bm E_{n_s}^{\Hh}\bm h_e$. Stacking the user channels gives
$\bar{\bm H}=[\bar{\bm h}_1,\ldots,\bar{\bm h}_K]^{\Hh}\in\mathbb C^{K\times n_s}$.
The BS jointly designs the port selection matrix $\bm E_{n_s}$ and the
beamforming matrix $\bm W=[\bm w_1,\ldots,\bm w_K]$, where
$\bm w_k\in\mathbb C^{n_s\times1}$ is the beamformer for legitimate user
$k\in\mathcal K$. Let $\bm x_d=[x_1,\ldots,x_K]^{\T}$ collect the unit-power
data symbols, i.e., $\mathbb E[|x_k|^2]=1$. The transmitted signal is expressed as 
\begin{equation}
    \bm x=\bm E_{n_s}\bm W\bm x_d
    =\sum_{k\in\mathcal K}\bm E_{n_s}\bm w_k x_k.
\end{equation}
Because the columns of $\bm E_{n_s}$ are orthonormal, the transmit-power
constraint becomes
\begin{equation}
    \|\bm E_{n_s}\bm W\|_F^2
    =\|\bm W\|_F^2
    =\sum_{k\in\mathcal K}\|\bm w_k\|_2^2
    \leq P_{\max}.\label{eq:power-constraint}
\end{equation}
The receiver noises satisfy $n_k\sim\CN(0,\sigma_k^2)$ and $n_e\sim
\CN(0,\sigma_e^2)$. The input-output relationship is given by 
\begin{subequations}
\begin{align}
  y_k&=\bm h_k^{\Hh}\bm E_{n_s}\bm w_kx_k+\sum_{j\ne k}\bm h_k^{\Hh}\bm E_{n_s}\bm w_jx_j+n_k,\\
 y_e&=\sum_{k\in\mathcal K}\bm h_e^{\Hh}\bm E_{n_s}\bm w_kx_k+n_e.
\end{align}
\end{subequations}
Under single-user decoding, the eavesdropper treats all other streams as
interference. The SINRs for user $k$ and eavesdropper are given by
\begin{subequations}
\begin{align}
 \gamma_k&=\frac{|\bm h_k^{\Hh}\bm E_{n_s}\bm w_k|^2}
 {\sum_{j\in\mathcal K\setminus\{k\}}|\bm h_k^{\Hh}\bm E_{n_s}\bm w_j|^2+\sigma_k^2},
 \quad k\in\mathcal K,\\
 \gamma_{e,k}&=\frac{|\bm h_e^{\Hh}\bm E_{n_s}\bm w_k|^2}
 {\sum_{j\in\mathcal K\setminus\{k\}}|\bm h_e^{\Hh}\bm E_{n_s}\bm w_j|^2+\sigma_e^2},
  \quad k\in\mathcal K. 
\end{align}
\end{subequations}
Accordingly, the achievable secure rate of user $k$ is
\begin{equation}
 R^s_k=\left[\log_2(1+\gamma_{k})-\log_2(1+\gamma_{e,k})\right]^+,  \quad k\in\mathcal K,
 \label{eq:secrecy-rate}
\end{equation}
following the standard wiretap-rate construction \cite{wyner1975wiretap}.

\subsection{Joint Secure Port Selection and Beamforming Problem}
Mathematically, the joint secure port selection and beamforming problem for
the multiuser downlink FAS is formulated as follows:
\begin{equation*}
\begin{aligned}
\mathrm{P1}:\quad
&\mathop{\mathrm{max}}_{\bm E_{n_s},\,\bm W}\quad
  \sum_{k \in \mathcal{K}}R^s_k(\bm E_{n_s},\bm W)\\
&\mathrm{s.t.}\quad \eqref{eq:port-membership},\
  \eqref{eq:port-distinct},\ \text{and}\ \eqref{eq:power-constraint}.
\end{aligned}
\end{equation*}
Physically, selecting a port changes the sampled channels of every user and
of the eavesdropper at the same time. Therefore, our aim is to select the ports that strengthen the desired links while suppressing multiuser interference and information leakage, subject to the total transmit power. Because port selection creates a combinatorial discrete space and the secrecy sum-rate is nonlinear in the selection and beamforming variables, P1 is a nonlinear mixed-integer nonconvex program.

\section{ReEvo-Designed Memetic Algorithm}
\begin{figure}[!t]
    \centering
    \includegraphics[width=0.925\columnwidth]{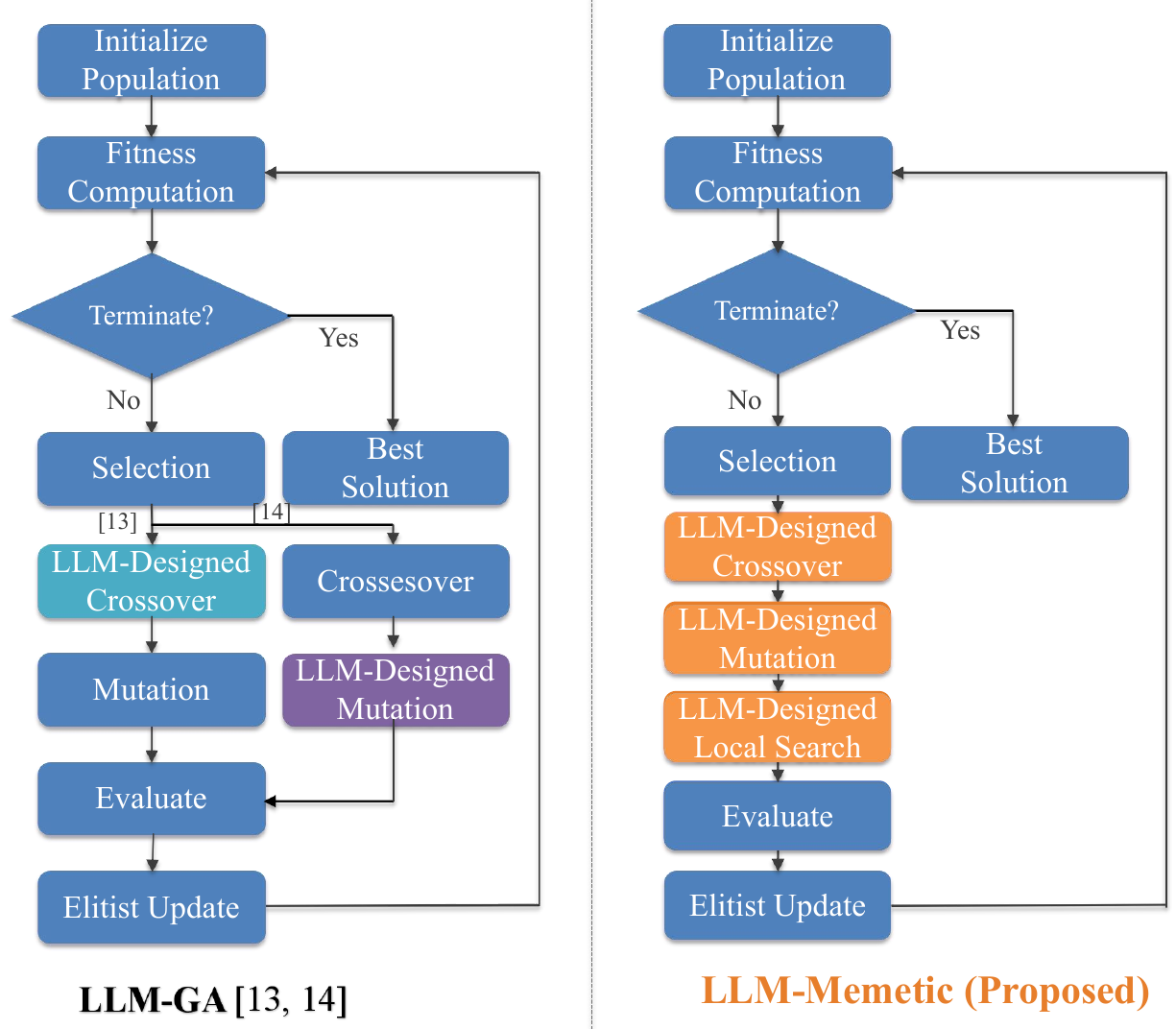}
    \caption{Flowchart comparison of \cite{wang2026fas} and \cite{zheng2026llmfas} with the proposed algorithm.}
    \label{fig:llm-memetic-flowchart}
\end{figure}

To solve P1, we develop a ReEvo-designed memetic algorithm comprising an offline LLM-driven operator design stage and an online memetic search stage that requires no LLM application programming interface (API) calls. In the offline stage, ReEvo evolves three complementary discrete operators, i.e., crossover, mutation, and local search, under a secrecy-driven fitness evaluated via secure WMMSE beamforming. In the online stage, each candidate port subset generated by the frozen operators is scored via secure WMMSE beamforming, and the memetic search then evolves toward higher secure sum-rates without any LLM queries. Fig.~\ref{fig:llm-memetic-flowchart} illustrates the overall architecture and highlights the distinction between our algorithm and prior LLM-GAs that design only a crossover operator \cite{wang2026fas} or only a mutation operator \cite{zheng2026llmfas}.

\subsection{Secure WMMSE Beamforming}
\setcounter{equation}{10}
For a fixed $\bm E_{n_s}$, we use a safeguarded WMMSE-based
procedure \cite{shi2011wmmse}.
\rev{Initialize $\bm V=\bar{\bm H}^{\Hh}
(\bar{\bm H}\bar{\bm H}^{\Hh}+10^{-6}\bm I)^{-1}$ and
$\bm W^{(0)}=\sqrt{P_{\max}}\bm V/\|\bm V\|_F$.}
With the scalar estimate $\hat{x}_k=u_k^*y_k$ at the current beamformer
$\bm W^{(t)}$, the user's minimum mean-square
error (MMSE) receiver, mean-square error (MSE), and weight are
\begin{subequations}\label{eq:legit-wmmse}
\begin{align}
T_k&=\sum_{j\in\mathcal K}|\bar{\bm h}_k^{\Hh}\bm w_j|^2+\sigma_k^2,\label{eq:legit-T}\\
u_k&=\frac{\bar{\bm h}_k^{\Hh}\bm w_k}{T_k},\qquad
\varepsilon_k=1-\frac{|\bar{\bm h}_k^{\Hh}\bm w_k|^2}{T_k},\label{eq:legit-u}\\
q_k&=\varepsilon_k^{-1}.\label{eq:legit-q}
\end{align}
\end{subequations}
When the eavesdropper decodes stream $k$ and treats the remaining streams as interference, its corresponding variables at the same $\bm W^{(t)}$ are given by
\begin{subequations}\label{eq:eve-wmmse}
\begin{align}
T_e&=\sum_{j\in\mathcal K}|\bar{\bm h}_e^{\Hh}\bm w_j|^2+\sigma_e^2,\label{eq:eve-T}\\
u_{e,k}&=\frac{\bar{\bm h}_e^{\Hh}\bm w_k}{T_e},\qquad
\varepsilon_{e,k}=1-\frac{|\bar{\bm h}_e^{\Hh}\bm w_k|^2}{T_e},\label{eq:eve-u}\\
q_{e,k}&=\varepsilon_{e,k}^{-1}.\label{eq:eve-q}
\end{align}
\end{subequations}
\begin{revision}
For an arbitrary receiver $u$, the MSE is
$e_k(u,\bm W)=|u|^2T_k$\allowbreak
${}-2\operatorname{Re}\{u^*\bar{\bm h}_k^{\Hh}\bm w_k\}+1$, so
$\ln(1+\gamma_k)=\max_{u\in\mathbb C,q>0}\{\ln q-qe_k+1\}$ and
$-\ln(1+\gamma_{e,k})=\min_{u\in\mathbb C,q>0}\{qe_{e,k}-\ln q-1\}$,
attained at \eqref{eq:legit-u}--\eqref{eq:legit-q} and
\eqref{eq:eve-u}--\eqref{eq:eve-q}, respectively.
The frozen auxiliary variables lower-bound the legitimate rate and
upper-bound the negative eavesdropper rate. The two bounds do not
combine, so the model below is only tangent to the unclipped rate
difference at $\bm W^{(t)}$ and its step need not improve the clipped
objective. Monotonicity follows from the line search below. Omitting the
common $1/\ln 2$ factor and the $\bm W$-independent constants gives
\end{revision}
\begin{equation}\label{eq:w-update}
\widetilde{\bm W}=\arg\max_{\|\bm W\|_F^2\le P_{\max}}
-\operatorname{tr}(\bm W^{\Hh}\bm A\bm W)
+2\operatorname{Re}\{\operatorname{tr}(\bm B^{\Hh}\bm W)\},
\end{equation}
where
\begin{subequations}\label{eq:matrix-updates}
\begin{align}
\bm A&=\sum_{k\in\mathcal K}q_k|u_k|^2
\bar{\bm h}_k\bar{\bm h}_k^{\Hh},\label{eq:a-update}\\
\bm b_k&=q_ku_k\bar{\bm h}_k
+c_e\bar{\bm h}_e\bar{\bm h}_e^{\Hh}\bm w_k^{(t)}
-q_{e,k}u_{e,k}\bar{\bm h}_e,\label{eq:b-update}\\
c_e&=\sum_{j\in\mathcal K}q_{e,j}|u_{e,j}|^2,\quad
\bm B=[\bm b_1,\ldots,\bm b_K].\label{eq:c-update}
\end{align}
\end{subequations}
\begin{revision}
The term $c_e\bar{\bm h}_e\bar{\bm h}_e^{\Hh}\bm w_k^{(t)}$ in
\eqref{eq:b-update} linearizes the leakage term
$\operatorname{tr}(\bm W^{\Hh}\bm C_e\bm W)$ of the surrogate, with
$\bm C_e=c_e\bar{\bm h}_e\bar{\bm h}_e^{\Hh}\succeq0$ convex in
$\bm W$: by convexity,
$\bm w_k^{\Hh}\bm C_e\bm w_k\ge
2\operatorname{Re}\{(\bm w_k^{(t)})^{\Hh}\bm C_e\bm w_k\}-
(\bm w_k^{(t)})^{\Hh}\bm C_e\bm w_k^{(t)}$,
so replacing this term by its tangent at $\bm W^{(t)}$ keeps
\eqref{eq:w-update} concave. Its Lagrangian stationarity condition is
$(\bm A+\mu\bm I)\widetilde{\bm W}=\bm B$, and hence
\end{revision}
\begin{equation}
\widetilde{\bm W}=(\bm A+\mu\bm I)^{-1}\bm B.
\end{equation}
\begin{revision}
Because $\bm A\succeq0$, the surrogate in \eqref{eq:w-update} is concave,
so the KKT conditions are sufficient: $\mu=0$ if the unconstrained
maximizer exists and is feasible. Otherwise the power constraint is
active with the unique $\mu>0$ satisfying
$\|\widetilde{\bm W}\|_F^2=P_{\max}$. A singular $\bm A$ may prevent
the first case, in which the second case applies.

A backtracking line search on the segment
$\bm W(\eta)=(1-\eta)\bm W^{(t)}+\eta\widetilde{\bm W}$,
$\eta\in(0,1]$, evaluates $\sum_{k\in\mathcal K}R_k^s$ via
\eqref{eq:secrecy-rate}. A negative per-user difference is clipped to
zero while still causing interference, and all streams are retained in
the unclipped model. The search accepts the largest nondecreasing step
and keeps the incumbent otherwise.
In exact arithmetic, monotone acceptance preserves feasibility and gives
\[
0\le\sum_{k\in\mathcal K}R_k^s\le\sum_k\log_2
\left(1+\frac{P_{\max}\|\bar{\bm h}_k\|_2^2}{\sigma_k^2}\right).
\]
Thus exact monotone objective values converge if continued indefinitely,
but this is not a beamformer or stationary-point convergence guarantee.
\end{revision}

\subsection{ReEvo-Designed Memetic Port Search}

We encode a port subset as a binary chromosome
$\bm s\in\{0,1\}^{N_s}$ satisfying $\bm 1^{\T}\bm s=n_s$. Following \cite{ye2024reevo}, the LLM acts as a hyper-heuristic. Let $\mathcal Q=\{c,m,l\}$ denote the crossover, mutation, and local search operator classes, respectively. The DeepSeek reasoning model evolves each class separately. For class $q$, each program $\chi_q$ is an individual in population $\Phi_q$. Let $\mathcal S_q(\chi_q)$ be the chromosomes visited by the inner memetic search that applies $\chi_q$ for class $q$ with the other operators fixed, on the training channels. The meta-fitness of $\chi_q$ is then
\begin{equation}
 F_q(\chi_q)=\max_{\bm s\in\mathcal S_q(\chi_q)}
 \sum_{k\in\mathcal K}R_k^{s}(\bm s). \label{eq:reevo-fitness}
\end{equation}
Since \eqref{eq:reevo-fitness} includes secure WMMSE optimization, programs are selected by the
target secrecy objective rather than a surrogate port score.

Crossover combines two parents using population fitness and ranks,
mutation performs a one-in--one-out swap, and local search generates
four neighbors: nearby, long-distance, same-row/column, and
crowded-to-sparse relocation.
Returned chromosomes are repaired to contain exactly $n_s$ active ports,
and invalid programs are discarded. All evaluations share the same
training channels, beamformer initialization, and stopping rule.

\begin{algorithm}[!t]
\caption{\normalcolor Offline LLM operator design and API-free secure FAS optimization}
\label{alg:reevo-memetic}
\begin{algorithmic}[1]
\FOR{$q\in\{c,m,l\}$ in order}
    \STATE Initialize program population $\Phi_q$ from seed and LLM code.
    \WHILE{fewer than $N_{\rm eval}$ programs have been evaluated}
        \STATE Evaluate training fitness $F_q$ by secure WMMSE.
        \STATE Reflect on higher/lower-fitness code, generate programs, and update $\Phi_q$.
    \ENDWHILE
    \STATE Freeze the best program $\chi_q^*$ by training fitness.
\ENDFOR
\STATE Initialize $P$ feasible chromosomes, then evaluate and cache their fitness and beamformers.
\FOR{$g=1,\ldots,G$}
    \STATE Preserve $E$ elites and select parents for $P-E$ children.
    \STATE Apply $\chi_c^*$ and $\chi_m^*$, then repair cardinality.
    \STATE Apply $\chi_l^*$ to create four neighbors per child.
    \STATE Score uncached candidates by secure WMMSE.
    \STATE Retain the best $P-E$ candidates with the $E$ elites.
\ENDFOR
\RETURN Best chromosome and its cached beamformer.
\end{algorithmic}
\end{algorithm}

Algorithm~\ref{alg:reevo-memetic} summarizes the procedure, in which
crossover is trained with conventional mutation and no local search,
mutation with the crossover fixed, and local search with both fixed.
For each class, the program with the best training fitness is then
checked on separate validation channels, and the three chosen programs
are frozen.
In the online stage, the frozen operators generate offspring and
local search neighbors, repeated chromosomes are scored once by secure
WMMSE beamforming, and elitist replacement keeps the $E$ elites and
the best $P-E$ candidates.

\section{Simulations}
Table~\ref{tab:sim-params} summarizes the simulation parameters.
The power sweep uses normalized gains $\beta_k=\beta_e=1$ and noise
variances $\sigma_k^2=\sigma_e^2=1$, i.e., $r_k=r_e=r_0$,
$\beta_0=1$, and the path-loss exponent cancels.  Each realization includes
all $K=10$ user channels and one eavesdropper channel. All methods share test channels, initial populations, and random seeds at each power. Offline ReEvo uses the DeepSeek reasoning model with a $10$-program population. Fitness uses $10$ training channels, and selected programs are checked on $20$ disjoint validation channels. Operator design uses $20$ training and $50$ validation generations at $P_{\max}=2.5$ W. Once three operators are designed, testing uses $20$ separate channels and $20$ random seeds. The online algorithm uses frozen code, with no neural-network training and no LLM calls.

\begin{figure}[!t]
    \centering
    \includegraphics[width=0.95\columnwidth]{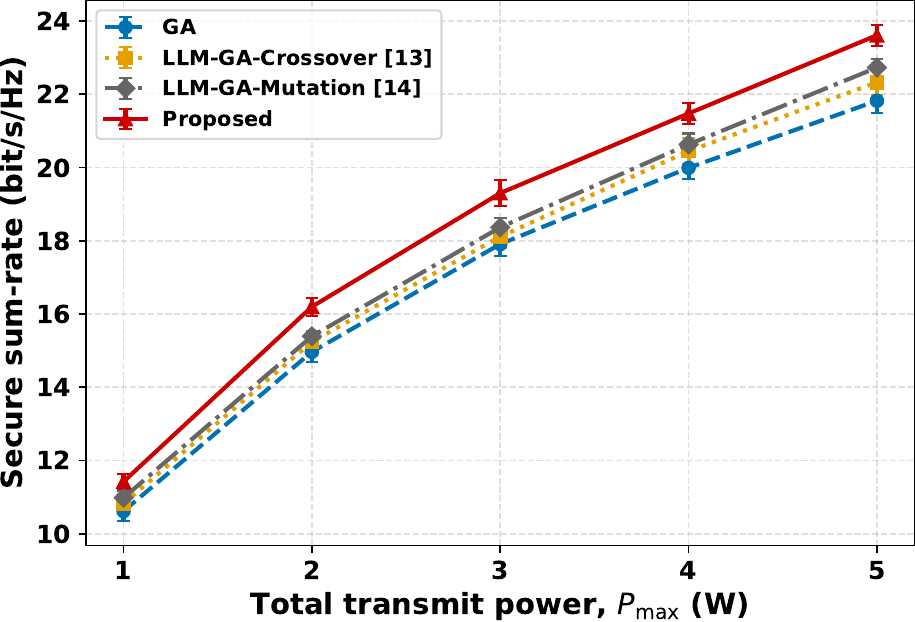}
    \caption{Mean secure sum-rate versus total transmit power when testing.}
    \label{fig:power-results}
\end{figure}

\begin{figure*}[!t]
    \setlength{\abovecaptionskip}{3pt}
    \centering
    \subfloat[1.0 W]{\includegraphics[width=0.2\textwidth]{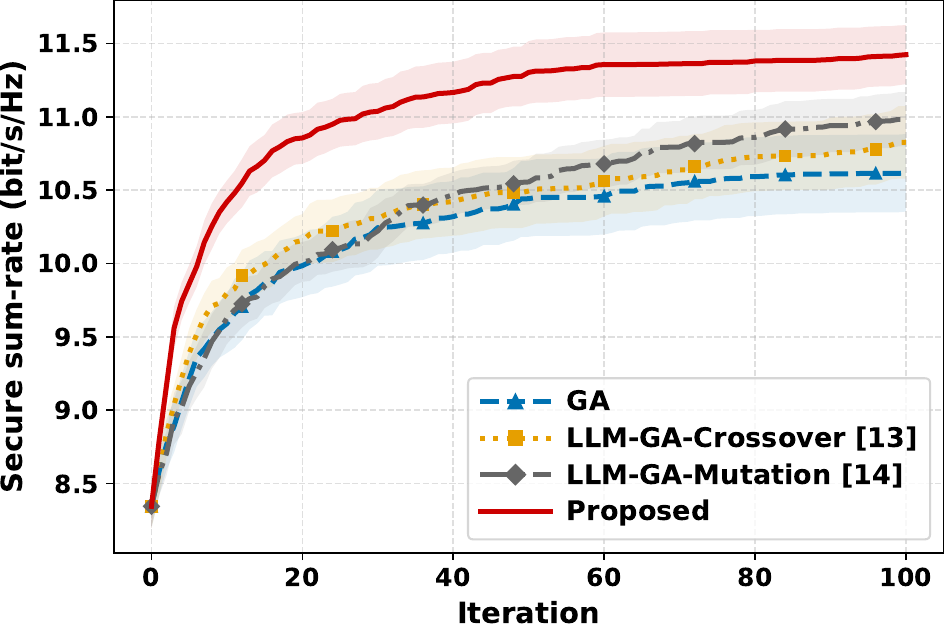}}%
    \subfloat[2.0 W]{\includegraphics[width=0.2\textwidth]{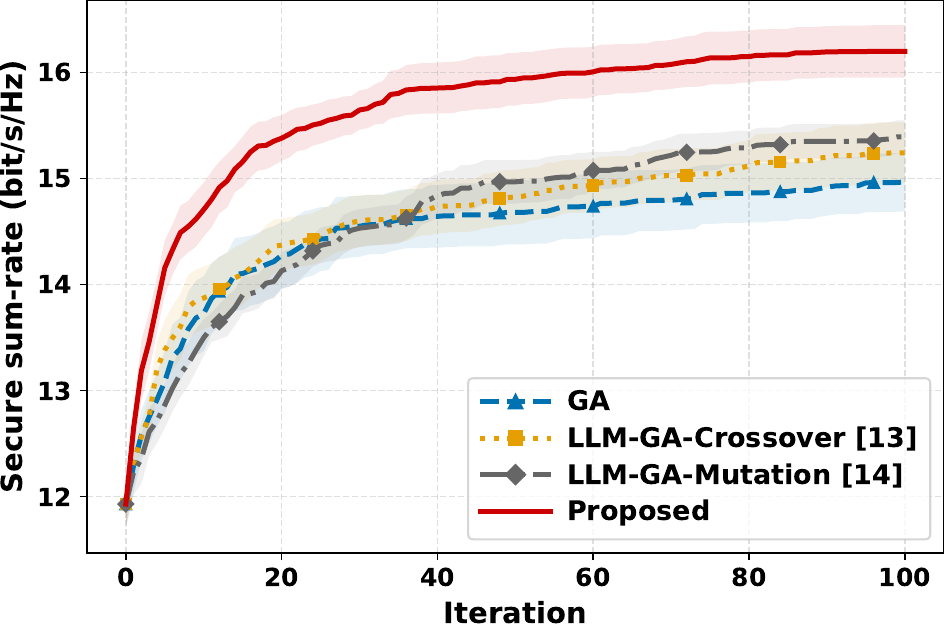}}%
    \subfloat[3.0 W]{\includegraphics[width=0.2\textwidth]{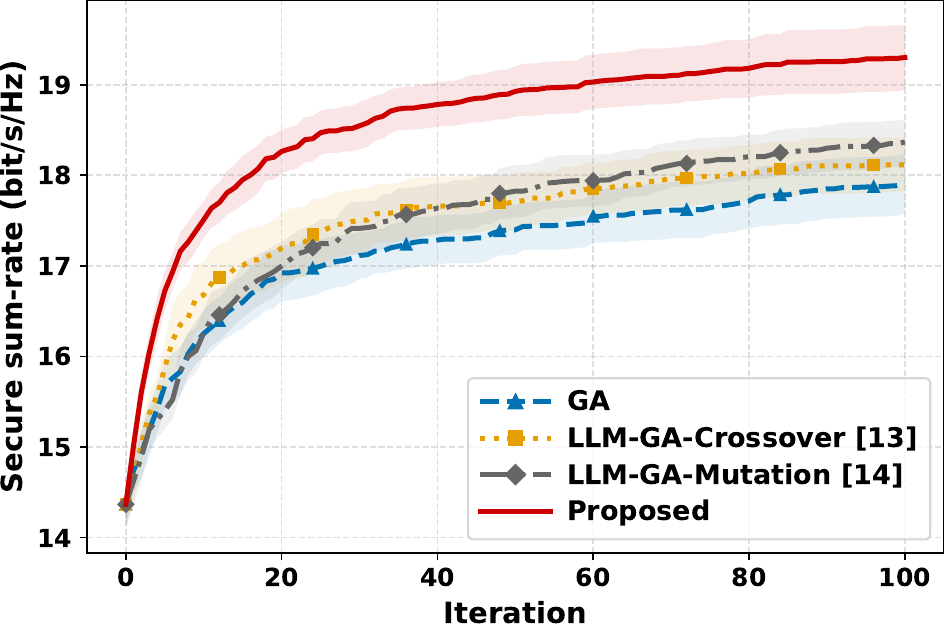}}%
    \subfloat[4.0 W]{\includegraphics[width=0.2\textwidth]{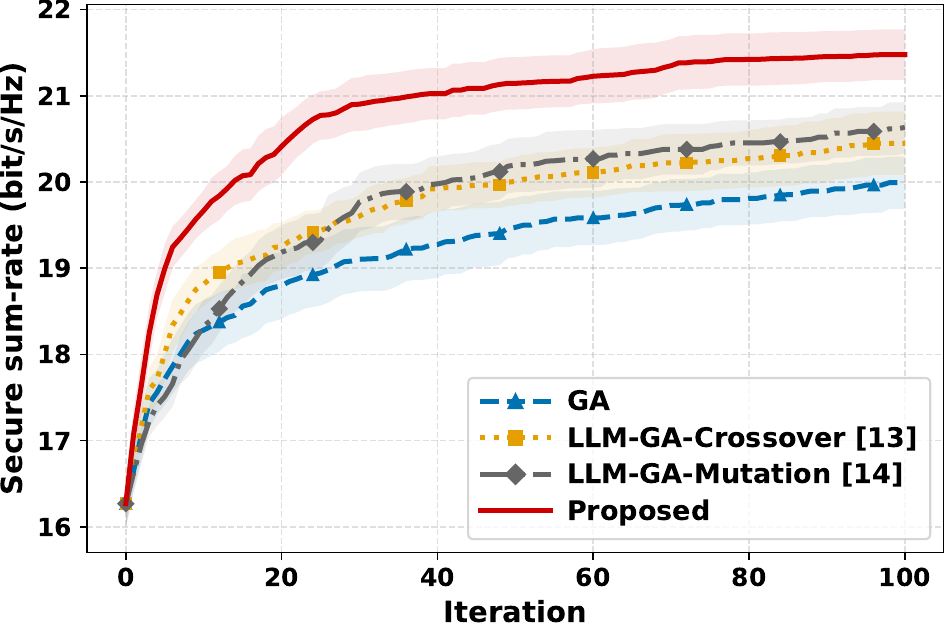}}%
    \subfloat[5.0 W]{\includegraphics[width=0.2\textwidth]{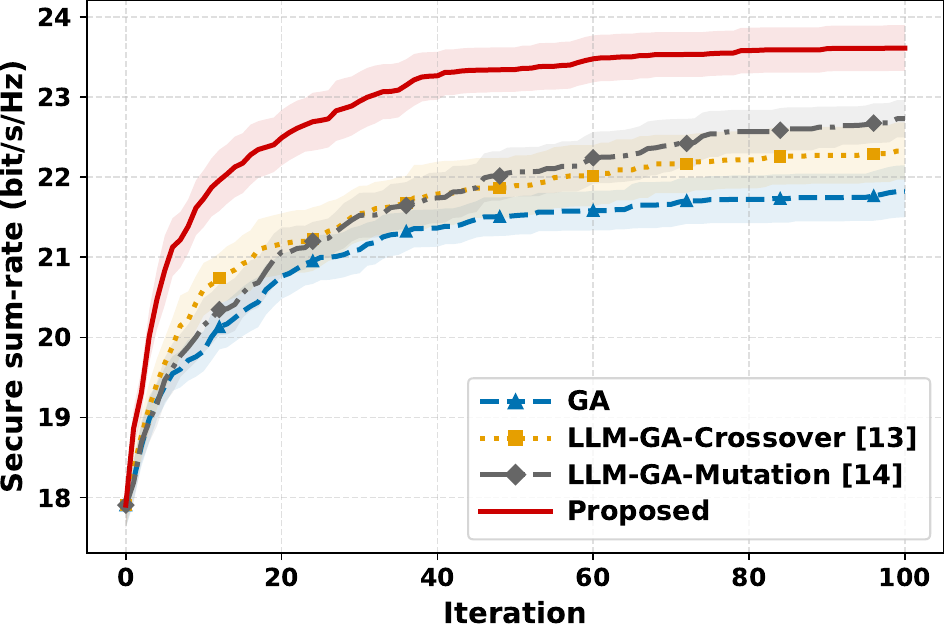}}
    \caption{Secure sum-rate convergence for five total transmit power levels when testing.}
    \label{fig:power-convergence}
\end{figure*}

\rev{Throughout, the total secure rate is summed over
users and averaged over test channels, never divided by $K$. The same test channels are reused across all total transmit power levels.} The error bars show how much the secure sum-rate varies across the 20 test channels, i.e., one standard deviation of the 20 per-channel values.

\begin{table}[!t]
\centering
\caption{Simulation parameters.}
\label{tab:sim-params}
%\footnotesize
\renewcommand{\arraystretch}{0.9}
\begin{tabular}{p{0.40\columnwidth}p{0.48\columnwidth}}
\hline
\rowcolor{gray!25} Parameter & Setting\\
\hline
Grid / aperture & $20\times20$ / $0.5\times0.5$~m$^2$\\
Carrier frequency & $3.4$~GHz\\
Users / eavesdropper & $K=10$ / 1\\
Active ports & $n_s=10$\\
\rev{Distances / path loss} & \rev{$r_k=r_e=r_0$, $\beta_k=\beta_e=1$}\\
\rev{Noise variance} & \rev{$\sigma_k^2=\sigma_e^2=1$ (normalized)}\\
Transmit power & $P_{\max}\in\{1,2,3,4,5\}$~W\\
Population / elites & $P=20$ / $E=4$\\
Offspring / generations & 16 / 100\\
Mutation probability & 0.2\\
\rev{Operator train / validation} & \rev{10 / 20 disjoint channels}\\
\rev{Test channels / random seeds} & \rev{20 paired / 20 seeds}\\
\rev{LLM / programs} & \rev{DeepSeek / 10-program population}\\
\hline
\end{tabular}
\end{table}

Fig.~\ref{fig:power-results} shows that the secure sum-rate of all four algorithms increases as $P_{\max}$ grows from $1$ to $5$ W when testing, with the proposed algorithm remaining the highest at every point. At 1 W, it achieves $11.423$~bit/s/Hz, corresponding to gains of $7.6\%$, approximately $6\%$, and approximately $4\%$ over GA, LLM-GA-crossover \cite{wang2026fas}, and LLM-GA-mutation \cite{zheng2026llmfas}, respectively. At 5 W, it achieves $23.611$~bit/s/Hz, yielding gains of $8.2\%$, approximately $5\%$, and approximately $4\%$ over the same three baselines. The two LLM-GA baselines consistently lie between GA and the proposed algorithm, with visibly overlapping error bars at some operating points. A possible explanation is that elite-aware recombination preserves useful spatial patterns, while local and global exchanges explore the tradeoffs among signal strength, interference, and leakage. This ordering holds at equal generation counts, though the additional local search evaluations may particularly account for the gains obtained by the proposed algorithm. Each operating point in Fig.~\ref{fig:power-results} is the generation-$100$ value of the corresponding curve in Fig.~\ref{fig:power-convergence}.

Fig.~\ref{fig:power-convergence} shows secure sum-rate convergence over $100$ generations (iterations) when testing. The proposed algorithm separates from the baselines early and stays ahead in all five power panels, while all curves gradually flatten as the populations approach stable incumbents. The terminal values at generation $100$ correspond to the respective operating-point results summarized in Fig.~\ref{fig:power-results}, linking the convergence endpoint to the power-sweep comparison. Local search evaluates additional one-swap neighbors, so these curves do not establish faster convergence in runtime or evaluation count.

\section{Conclusion}
This paper presented a ReEvo-designed memetic algorithm for joint port selection and secure beamforming in multiuser fluid antenna systems. An LLM evolves dedicated crossover, mutation, and local search operators offline under a secrecy-driven fitness, and the frozen programs incur no online LLM-query overhead. For each candidate port configuration, the beamformer is optimized by secure WMMSE.

Across the tested total transmit power levels, the proposed method consistently obtains the highest mean secure sum-rate among the conventional GA and both LLM-GA baselines, with gains that remain visible from low to high transmit power. These results indicate that evolving all three discrete search operators offline can outperform LLM-GAs that design only a single operator. Extending the framework to imperfect or delayed CSI and to more general multiuser and multi-antenna scenarios is left as future work.

\end{document}